\documentclass[twocolumn,showpacs,preprintnumbers, prl,
               superscriptaddress]{revtex4}
\usepackage{graphicx, fancybox}
\usepackage{amsmath,amssymb}

\newcommand{\bm}[1]{\mbox{\boldmath$#1$}}

\begin{document}

\preprint{BNL-NT-07/34}

\title{
Spectral properties of quarks above $T_c$
in quenched lattice QCD
}

\author{Frithjof Karsch}
\email{karsch@quark.phy.bnl.gov}
\affiliation{
Brookhaven National Laboratory, Bldg.510A, Upton, 11973, NY, USA}

\author{Masakiyo Kitazawa}
\email{kitazawa@phys.sci.osaka-u.ac.jp}
\affiliation{
Department of Physics, Osaka University, Toyonaka, Osaka, 560-0043, Japan
}

\begin{abstract}

We analyze the quark spectral function
above the critical temperature for deconfinement
in quenched lattice QCD using clover improved
Wilson fermions in Landau gauge.
We show that the temporal quark correlator is well
reproduced by a two-pole approximation for the spectral
function and analyze the bare quark mass dependence
of both poles as well as their residues.
In the chiral limit we find that the quark spectral function
has two collective modes which correspond to the normal and
plasmino excitations. At large values of the bare quark mass
the spectral function is dominated by a single pole.

\end{abstract}

\date{August 2, 2007}

\maketitle


To explore the properties of hot and dense matter
formed by quarks and gluons above the critical temperature 
for deconfinement ($T_c$) is an intriguing problem that has
been addressed in many studies. Recent 
experimental results on the properties of the matter 
created in heavy ion collisions at 
the Relativistic Heavy-Ion Collider (RHIC) suggest, that 
its time evolution above $T_c$ is well described by ideal 
hydrodynamics down to the freeze-out temperature in the 
vicinity of $T_c$ \cite{RHIC}.
In order to understand better
the structure of matter in this non-perturbative region, 
it is desirable to identify the basic degrees of freedom 
of the system and their quasi-particle properties. 

At asymptotically high temperatures almost free
quarks and gluons are most certainly the basic
degrees of freedom that control the properties of
the Quark-Gluon plasma (QGP) \cite{LeBellac}. In
this regime properties of the QGP can be analyzed using 
perturbative techniques. At lower temperatures the
application of hard-thermal loop (HTL) resummation
\cite{HTL} still allowed to define
gauge independent propagators for quarks and gluons
that can be used to study properties of the QGP 
perturbatively. From these perturbative analyses it
is known that the collective excitations of gluons and quarks  
develop a mass gap (thermal mass) that is proportional 
to $gT$ \cite{plasmino,HTL,LeBellac}, where $g$ and $T$ denote 
the gauge coupling and temperature, respectively.
Moreover, the number of poles in the finite temperature 
quasi-particle propagators is doubled. 
In addition to the normal modes,
which reduce to poles in the free particle propagator, 
plasmon and plasmino modes appear.

At temperatures in the vicinity of $T_c$ it is apriori not clear 
whether a quasi-particle picture for quarks and gluons is valid 
at all. However, lattice results on e.g. baryon number and electric
charge fluctuations in the vicinity of $T_c$ \cite{fluctuations} suggest 
that quasi-particles with quark degrees of freedom are the carriers 
of these quantum numbers. Quasi-particles also have been used
successfully to describe lattice QCD results on the equation
of state \cite{EoS}. Moreover, the
apparent quark number scaling of the elliptic flow observed in 
the RHIC experiments \cite{Fries:2003kq} may also suggest  
that quasi-particles with quark quantum numbers exist even 
close to $T_c$.
Despite the problem of gauge dependence of quark and gluon propagators, 
it therefore is desirable to analyze their properties at high 
temperature through a direct calculation within the framework of QCD. 
In this Letter, we analyze dynamical properties of 
quarks above  $T_c$ in quenched lattice QCD. These calculations
have been performed in Landau gauge. 
So far, there have been only a few studies that address this problem
in lattice calculations \cite{quark}.

In order to understand the origin of the plasmino mode in the quark
propagator in the high temperature limit,
it is instructive to consider the quark propagator 
at intermediate temperature by introducing some energy scale
of the order of the temperature
\cite{BBS92,Boyanovsky:2005hk,KKN06}.
In \cite{BBS92}, the temperature dependence of the 
spectral function for fermions with scalar mass $m$ 
has been considered in QED.
In this model, the spectral function at zero temperature 
has two poles at energies $\omega=\pm m$,
while in the high temperature limit, $T/m\to\infty$,
it approaches the HTL result, which has four poles.
The one-loop calculation performed in \cite{BBS92} clearly showed
that the two limiting forms of the spectral function are 
connected continuously;
in the spectral function a peak 
corresponding to the plasmino {gradually appears and
becomes larger } with increasing temperature, 
in addition to the normal quasi-particle peak \cite{BBS92}.

In this Letter, we analyze the quark propagator at two values of
the temperature, $T=1.5T_c$ and $3T_c$, as a function of the bare 
quark mass.
To simplify the present analysis, all our calculations have been
performed for zero momentum.
The dynamical properties of quarks at zero momentum are 
encoded in the quark spectral function $\rho( \omega )$
which is related to the Euclidean correlation function
\begin{align}
S( \tau ) 
= \frac1V \int d^3 x d^3 y 
\langle \psi( \tau,\bm{x} ) \bar\psi( 0,\bm{y} ) \rangle,
\label{eq:S}
\end{align}
through an integral equation
\begin{align}
S( \tau ) = \int_{-\infty}^{\infty}
d\omega \frac{ e^{ (1/2-\tau T)\; \omega/T } }
{ e^{\omega/2T} + e^{-\omega/2T} } \rho( \omega ),
\label{eq:SKrho}
\end{align}
with the quark field $\psi$, the spatial volume $V$, and
the imaginary time $\tau$ which is restricted to the interval 
$0<\tau<1/T$.
The Dirac structure of $\rho( \omega )$ is decomposed as
\begin{align}
\rho( \omega )
&= \rho_{\rm 0} ( \omega ) \gamma^0 + \rho_{\rm s} ( \omega )
\nonumber \\
&=\rho_+ ( \omega ) \Lambda_+ \gamma^0 
+ \rho_- ( \omega ) \Lambda_- \gamma^0,
\label{eq:rho}
\end{align}
with projection operators $ \Lambda_\pm = ( 1 \pm \gamma^0 )/2 $.
The charge conjugation symmetry leads to 
$\rho_{\rm 0} ( \omega ) =  \rho_{\rm 0} ( -\omega )$,
$\rho_{\rm s} ( \omega ) =  -\rho_{\rm s} ( -\omega )$, and
$\rho_+ (\omega) = \rho_- (-\omega)
 = \rho_{\rm 0} (\omega) + \rho_{\rm s} (\omega)$
\cite{BBS92,Weldon:1999th}. 
In the following analysis, we concentrate on a determination of 
$\rho_\pm(\omega)$ instead of $\rho_{\rm 0,s}(\omega)$,
because excitation properties of quarks are more apparent
in these channels \cite{BBS92}. 
In analogy to Eq.~(\ref{eq:rho}) we introduce the decomposition of the
correlation function $S( \tau )$ as $S (\tau)
=S_+ (\tau) \Lambda_+ \gamma^0 + S_- (\tau) \Lambda_- \gamma^0$, where
$S_\pm$ are related through $S_+(\tau) = S_-(\beta-\tau)$.

For free quarks with scalar mass $m_q$ the spectral functions,
$\rho_\pm (\omega) = \delta (\omega \mp m_q)$, 
have quark and anti-quark poles 
at $\omega= \pm m_q$, respectively.  
In the high temperature limit, additional poles,
corresponding to the plasmino, appear at negative energy for 
$\rho_+ (\omega)$ and positive energy for $\rho_- (\omega)$ 
\cite{BBS92,Weldon:1999th,KKN06}.
While the positivity of $\rho_\pm(\omega)$ is ensured by
definition, these spectral functions 
are neither even nor odd functions. In the chiral limit, however,
$\rho_{\rm s}$ vanishes and $\rho_\pm(\omega)$ become even functions.

To extract the spectral function $\rho_+(\omega)$ from $S(\tau)$
using Eq.~(\ref{eq:SKrho}), we assume that $\rho_+(\omega)$ 
can be described by a two-pole ansatz,
\begin{align}
\rho_+(\omega)
= Z_1 \delta( \omega - E_1 ) + Z_2 \delta( \omega + E_2 ),
\label{eq:2pole}
\end{align}
where the residues $Z_{1,2}$ and energies $E_{1,2}>0$ have to be
determined from a fit to $S_+(\tau)$. 
The poles at $\omega=E_1, -E_2$ correspond to
the normal and plasmino modes, respectively
\cite{BBS92}.

\begin{table}
\begin{center}
\begin{tabular}{ccrcccc}
\hline
\hline
$T/T_c$ & $N_\tau$ & $N_\sigma$ & $\beta$ & $c_{\rm SW}$ & $\kappa_c$ & $a$[fm] \\
\hline
$3$     &    $16$ & $64,48$     & $7.457$ & $1.3389$     & $0.13390$  & $0.015$ \\
        &    $12$ & $48$        & $7.192$ & $1.3550$     & $0.13437$  & $0.021$ \\
$1.5$   &    $16$ & $64,48$     & $6.872$ & $1.4125$     & $0.13495$  & $0.031$ \\
        &    $12$ & $48$        & $6.640$ & $1.4579$     & $0.13536$  & $0.041$ \\
\hline
\hline
\end{tabular}
\end{center}
\caption{
Simulation parameters \cite{params}.
}
\label{table}
\end{table}

The correlation function $S(\tau)$ has been calculated at two
values of the temperature, $T =1.5T_c$ and $3T_c$, in 
quenched QCD using non-perturbatively improved clover Wilson 
fermions \cite{Sheikholeslami:1985ij,Luscher:1996jn}.
To control the dependence of our results on the finite lattice 
volume, $N_\sigma^3 \times N_\tau$, and lattice spacing, $a$,
we analyze the quark propagator on lattices of three different 
sizes. 
The gauge field ensembles used for this analysis have been
generated and used previously by the Bielefeld group to study
screening masses and spectral functions  \cite{params}.
The different simulation
parameters are summarized in Table~\ref{table} \cite{params}.
For each lattice size, $51$ configurations have been analyzed. 
On the $64^3\times 16$ lattices and at our smallest temperature,
$T=1.5T_c$,
we observed for the largest values of the hopping parameter, i.e.
closest to $\kappa_c$, an anomalous behavior of the quark propagator 
on a few gauge field configurations. The appearance of such exceptional
configurations in calculations with light quarks in quenched QCD
is a well-known problem in calulations with Wilson fermions 
\cite{excp.conf}.
We identified $7$ such configurations, which we excluded 
from our analysis.
The properties of the quark propagator on these configurations will
be discussed in more detail elsewhere \cite{KKprep}.
Quark propagators have been calculated after fixing each 
gauge field configuration to Landau gauge. For this we used
a conventional minimization algorithm with a stopping 
criterion, $(1/3){\rm tr}|\partial_\mu A^\mu|^2 <10^{-11}$.
In the Wilson fermion formulation the 
bare mass, $m_0$, is related to the  hopping parameter $\kappa$,
through the standard relation
\begin{align}
m_0 = \frac1{2a} \left( \frac1\kappa - \frac1{\kappa_c} \right) \; ,
\label{eq:m_0}
\end{align}
where $\kappa_c$ denotes the critical hopping parameter corresponding
to the chiral limit, or vanishing quark mass.

To evaluate Eq.~(\ref{eq:S}) numerically, 
we solve the linear equation
$K \psi_{\rm result} = \psi_{\rm source}$
for a given source $\psi_{\rm source}$,
with $K$ being the fermion matrix.
For this procedure, we use the wall source, 
$\psi_{\rm source}^{\rm w} = (1/V) \sum_{x} \psi( 0,\bm{x} )$,
which we found to be very efficient in reducing the
statistical error in the propagator calculation.
To reduce the statistical error further,
we define the correlation function $S_+(\tau)$
for each configuration by
\begin{align}
S_+^{\rm latt.}(\tau) 
= \frac1{12} {\rm tr} [S(\tau) \gamma^0 \Lambda_+
+ S(\beta-\tau) \gamma^0 \Lambda_-],
\label{eq:S_latt}
\end{align}
with the trace taken over Dirac and color indices.

\begin{figure}[tbp]
\begin{center}
\includegraphics[width=.49\textwidth]{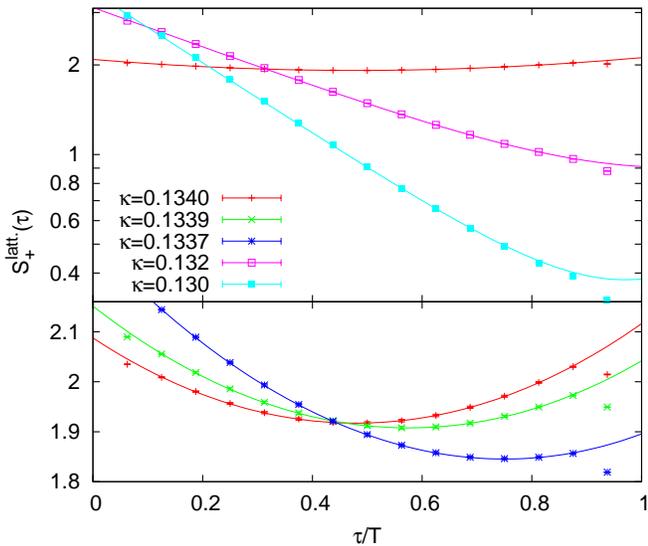}
\caption{
The lattice correlation function $S_+^{\rm latt.}(\tau)$
at $T=3T_c$ for the lattice of size $64^3\times16$
with various values of $\kappa$, and the 
fitting result with the ansatz Eq.~(\ref{eq:2pole}).
}
\label{fig:S}
\end{center}
\end{figure}

In Fig.~\ref{fig:S}, we show the numerical results
for $S_+^{\rm latt.}(\tau)$ for several values of $\kappa$
calculated on a lattice of 
size $64^3\times16$ at $T=3T_c$.
One sees that the shape of $S_+^{\rm latt.}(\tau)$
approaches that of a single exponential function for smaller $\kappa$,
while it becomes symmetric as $\kappa$ approaches $\kappa_c$.
In the vicinity of the wall source, {\it i.e.} at small and large
$\tau$, we see deviations from this generic picture which can
be attributed to distortion effects arising from the presence of the
source.
We thus exclude points with $\tau < \tau_{min}$ and
$N_\tau -\tau < \tau_{min}$ from our fits to the ansatz
given in Eq.~(\ref{eq:2pole}).
The resulting correlation functions obtained from correlated fits
with $\tau_{min}=3$ are shown in Fig.~\ref{fig:S}.
One sees that $S_+^{\rm latt.}$ is well reproduced by
our fitting ansatz \footnote{We also checked that fits 
based only on a single pole ansatz lead to 
unacceptable large  $\chi^2/{\rm dof}$.};
the $\chi^2/{\rm dof}$
of our fits is between $2$ and $3$ at 
$0.1335\lesssim\kappa\lesssim0.134$, 
while it gradually increases as $\kappa$ becomes smaller 
than $\kappa=0.1335$.
A similar behavior is also observed for our other lattice sizes \cite{KKprep}.

\begin{figure}[tbp]
\begin{center}
\includegraphics[width=.49\textwidth]{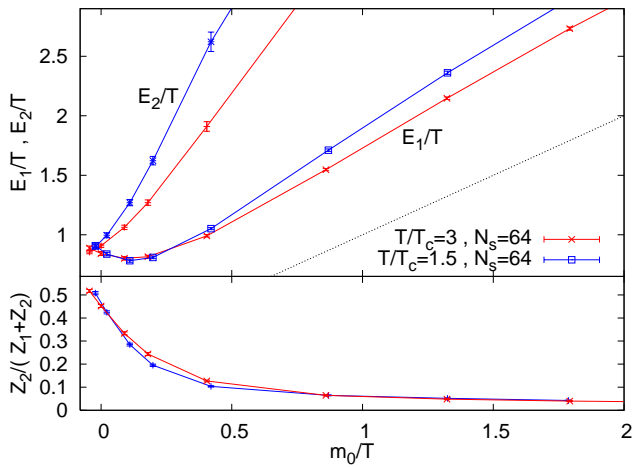}
\caption{
The bare quark mass dependence of fitting parameters 
$E_{1,2}$ and $Z_2 / ( Z_1+Z_2 )$ 
at $T=1.5T_c$ and $3T_c$ for lattice $64^3\times16$.
}
\label{fig:ZE}
\end{center}
\end{figure}

In Fig.~\ref{fig:ZE}, we show the dependence of 
$E_{1,2}$ and $Z_2 / ( Z_1+Z_2 )$ on the bare quark
mass $m_0$ for $T=1.5T_c$ and $3T_c$. 
The results have been obtained
from two-pole fits on lattices of size $64^3\times16$.
Errorbars have been estimated from a  Jackknife analysis.
The dotted line in this figure denotes the pole mass determined
from the bare lattice mass given in Eq.~(\ref{eq:m_0}), 
{\it i.e.} $E_1/T = m_0/T= m_0aN_\tau$.
The figure shows that
the ratio $Z_2 / (Z_1+Z_2)$ becomes larger with
decreasing $m_0$ and eventually reaches $0.5$.
The hopping parameters satisfying $Z_1=Z_2$ are
$\kappa'_c = 0.133974(10)$ for $T=3T_c$ and
$\kappa'_c = 0.134991(9)$ for $T=1.5T_c$, 
which are consistent with the values for
$\kappa_c$ given in Table~\ref{table}.
The latter had been obtained in \cite{params} from a fit 
to critical hopping parameters determined in \cite{Luscher:1996jn} 
from the vanishing of the isovector axial current.
The numerical results obtained on $64^3\times16$ lattices 
show that $E_1$ and $E_2$ are equal 
within statistical errors at $\kappa=\kappa'_c$.
The spectral function $\rho_+(\omega)$ thus becomes an 
even function at this point; the quark propagator becomes
chirally symmetric despite the presence of a thermal mass,
$m_T \equiv E_1=E_2$.
From Fig.~\ref{fig:ZE}, one also finds that the ratio $m_T/T$
is insensitive to $T$ in the temperature range analyzed in this work,
while it is slightly larger for lower $T$.

As $m_0$ becomes larger, $Z_2/(Z_1+Z_2)$ decreases and 
$\rho_+(\omega)$ is eventually dominated by a single-pole.
One sees that $E_1$ has a minimum at $m_0>0$,
while $E_2$ is an increasing function of $m_0$.
In the one-loop approximation, 
the peak in $\rho_+(\omega)$ corresponding to $E_1$ ($E_2$) is
monotonically
increasing (decreasing) function of $m_0/T$ \cite{BBS92,KKprep}.
The quark mass dependence of poles found here thus is
qualitatively different from the perturbative result.
We find, however, that slope of $E_2$ as function of $m_0/T$
decreases with increasing $T$. This may suggest that the
perturbative behavior could eventually be recovered at
much larger temperatures.

\begin{figure}[tbp]
\begin{center}
\includegraphics[width=.49\textwidth]{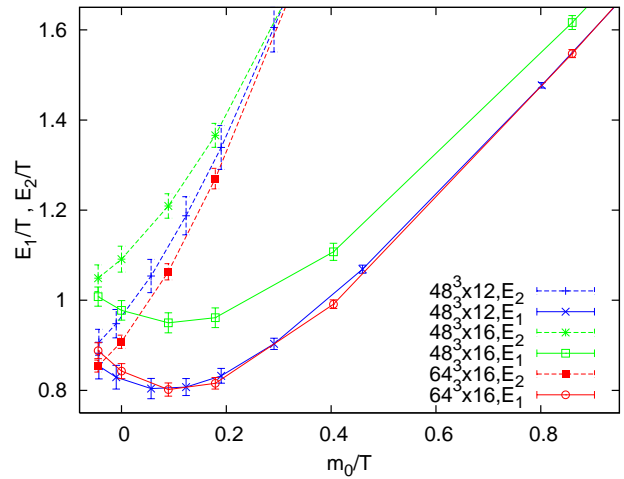}
\caption{
The bare quark mass dependence of parameters $E_1$, $E_2$ 
at $T=3T_c$ for lattices of size
$64^3\times16$, $48^3\times16$ and $48^3\times12$.
}
\label{fig:compare}
\end{center}
\end{figure}

In order to check the dependence of our results on the lattice 
spacing and finite volume, we analyzed the quark propagator 
at $T=3T_c$ for three different lattice sizes. Results for
$E_1$ and $E_2$ are shown in Fig.~\ref{fig:compare}. 
Comparing the results obtained on lattices with different 
lattice cut-off, $a$,
but same physical volume, {\it i.e.} $64^3\times16$ and 
$48^3\times12$, one sees that any possible cut-off dependence 
is statistically not significant in our analysis. On the other
hand we find a clear dependence of the quark 
energy levels on the spatial volume; when comparing lattices
with aspect ratio $N_\sigma/N_\tau =3$ and $4$ we find that the
energy levels, $E_{1,2}$, drop significantly. 
A similar behavior is observed also at $T=1.5T_c$.

The presence of a strong volume dependence of the quark propagator
is not unexpected. In fact, the thermal quark mass arises as 
collective effect of low momentum gluons; gluons at the soft 
scale $p\lesssim gT$ play a crucial role to give rise to the thermal 
mass at high temperatures \cite{HTL,LeBellac}. However, on lattices 
with given aspect ratio $N_\sigma/N_\tau$ low momentum gluons are 
cut-off. The lowest non-vanishing gluon momentum is,
$p_{min}/T = 2\pi (N_\tau / N_\sigma)$, which still is larger than
unity on lattices with aspect ratio $N_\sigma/N_\tau=4$. 
The situation may, nonetheless, be somewhat better in the
temperature range explored here as the temperature 
dependent coupling $g(T)$ is larger than unity.
An analysis of quark spectral functions on lattices
with even larger spatial volume is needed in the future to
properly control effects of small momenta. 
We attempted to estimate the thermal mass in the $V\to\infty$ limit
by extrapolating the results obtained for two different volumina.
Defining $m_T \equiv ( Z_1 E_1 + Z_2 E_2 ) / ( Z_1+Z_2) 
|_{\kappa = \kappa_c}$ and assuming
the volume dependence of $m_T$ as 
$m_T(N_\tau/N_\sigma)= m_T(0) \exp(c N_\tau^3/N_\sigma^3)$,
we obtain $m_T(0)/T = 0.771(18)$ for $T=3T_c$
and $m_T(0)/T = 0.800(15)$ for $T=1.5T_c$. This suggests that
finite volume effects may still be of the order of 15\% in
our current analysis of $m_T/T$. 
Despite these problems, our result clearly shows that
light quarks near but above $T_c$ have a mass gap that is 
of collective nature similar to that in the perturbative regime.

In this Letter,
we analyzed the quark spectral function at zero momentum 
for $T=1.5T_c$ and $3T_c$ as functions of bare quark mass $m_0$ 
in quenched lattice QCD with Landau gauge fixing.
We found that the two-pole approximation for $\rho_+(\omega)$
well reproduces the behavior of the lattice correlation function.
It is argued that the chiral symmetry of the quark propagator is
restored at the critical value of $\kappa$ and
the shape of the spectral function at this point takes
a similar form as in the high temperature limit
having normal and plasmino modes with thermal mass $m_T$.
As $m_0$ is increased, 
$\rho_+(\omega)$ approaches a single-pole structure
as one can naturally deduce intuitively.
The non-perturbative nature of thermal gauge fields is reflected 
in the behavior of poles as functions of $m_0$, which is 
qualitatively different from the perturbative result 
\cite{BBS92}.
We also note that the ratio $m_T/T$ decreases slightly
with increasing $T$, which is expected to happen at
high temperature where $m/T$ should be proportional to 
a running coupling $g(T)$.
Although results on the quark propagator are gauge dependent,
we expect that our results for its poles suffer less from gauge 
dependence,
because the success of the pole approximation for $\rho_+(\omega)$
indicates that the quark propagator 
has dynamical poles near the real axis, which are 
gauge independent quantities \cite{BBS92,g-dep}.

In the present study, we analyzed the quark spectral function
in the quenched approximation.
Although this approximation includes the leading contribution
in the high temperature limit \cite{LeBellac} and thus is valid 
at sufficiently high $T$, 
the validity of this approximation near $T_c$ is nontrivial.
For example, screening of gluons due to the 
polarization of the vacuum with virtual quark antiquark pairs
is neglected in this approximation.
The coupling to possible mesonic excitations \cite{J/y,HK85},
which may cause interesting effects in the spectral properties
of the quark \cite{KKN06}, are not incorporated, either.
The comparison of the quark propagator between quenched and
full lattice simulations would tell us 
the strength of these effects near $T_c$.

In the future it will also be interesting to use results on
the non-perturbative structure of quark propagators as input
for phenomenological studies of the QGP phase.
For example, thermal properties of the charm quark \cite{KKprep} 
should be useful for the understanding of properties of 
charmonia above $T_c$ \cite{J/y}.
The thermal mass of light quarks can also be used to
evaluate details of their dynamics \cite{Hidaka:2006gd}.

Although in this letter we limited our analysis to zero momentum,
a proper analysis at finite momentum \cite{KKprep} is needed 
for the understanding of the entire quark spectral function. 
In particular, the confirmation of the existence of a minimum in
the plasmino dispersion relation at finite momentum \cite{plasmino} 
clearly is a challenging problem.
The exploration of the gluon propagator is also an important
subject of further studies.
To clarify the origin of the quark mass dependence of $E_1$ and $E_2$, 
which is qualitatively different from the perturbative result,
as well as the $T$ dependence of the thermal mass,
are open questions for further numerical and analytic studies.


M.~K. is grateful to S.~Datta, W.~Soeldner and T.~Umeda
for helping in getting started with his first lattice simulation.
He also thanks S.~Ejiri for discussions.
The lattice simulations presented in this work 
have been carried out using the cluster computers 
ARMINIUS@Paderborn, BEN@ECT* and BAM@Bielefeld.
In the early phase of this project
M.~K. has been supported by Special Postdoctoral Research Program
of RIKEN.
F.~K. has been supported by contract DE-AC02-98CH10886 
with the U.S. Department of Energy.

\end{document}